\documentclass{article}

\usepackage{graphicx} 
\usepackage[utf8]{inputenc}
\usepackage{physics}
\usepackage{graphicx} 
\usepackage{epsfig}

\usepackage[english]{babel}

\usepackage[letterpaper,top=2cm,bottom=2cm,left=3cm,right=3cm,marginparwidth=1.75cm]{geometry}

\usepackage{amsmath}
\usepackage{url}

\usepackage[colorlinks=true, allcolors=blue]{hyperref}

\title{Women in STEM: Interview with Halina Abramowicz}
\author{Elisabetta Gallo$^{1,2}$, Henriette Ullmann$^{2}$ \\
$^1$Deutsches Elektronen-Synchroton DESY, 
Notkestr. 85, 22607 Hamburg, Germany \\
$^2$Universit{\"a}t Hamburg, Germany}

\usepackage[style=numeric,sorting=none]{biblatex}
\usepackage{comment}
\begin{document}

\maketitle
\tableofcontents

\begin{abstract}
    This short article is a first of a series  describing the scientific journey of exceptional women scientists in
experimental particle physics. We interviewed Halina Abramowicz, who started her career in hadron-hadron interactions, in neutrino physics, became an expert of strong interactions, guided the
European Particle Physics Strategy Update in 2020 and now moved to an experiment in strong-field QED.
\end{abstract}

\section{Introduction}

In the scientific career of many women colleagues in experimental particle physics, we have met exceptional women, whose  difficulties at the beginning of their scientific path and later achievements are not very well known. In this short preprint, we interview Halina Abramowicz of Tel Aviv University, to learn about her journey through hadron interactions and deep inelastic scattering with neutrinos and charged leptons, from her initial studies in Warsaw, to her experience at CERN and DESY and her contributions to the future of our field.

\section{Rational choice}

Halina Abramowicz performed all her studies in Warsaw, Poland, covering a period of about 15 years from her master to her habilitation. While at the master level only 5\% of the students  attending the courses were women, a higher percentage continued in research, a sign of the fact that if women were studying, they had a real dedication to the field. The economical situation of Poland at that time required both women and men in a family to work, so that women chose also professional careers typical of men.  Even if the head of the group was a man, there were actually many women in the high energy physics department in Warsaw at that time, in the 70s, likely due to the aftermath of World War II.

There was  only one occasion when Halina was made aware that there was an issue between men and women in physics. Being more inclined to theoretical particle physics,  she asked a master thesis topic in theory. She was then told that she should better choose a master topic in experimental particle physics as, if there would have been a choice between a woman or a man for a staff position in theory, a man would be selected. 

Therefore, she made the rational choice to move to experimental particle physics, with a master thesis on $\pi^0$ studies with the 2\,m bubble chamber at CERN. It was the time when people were looking into inclusive production to understand the dynamics of the strong interactions. She was part of the scanning, reconstruction and analysis of $\pi^0$ mesons~\cite{Bockmann:1976dh}
through their decay into two photons converted in the bubble chamber. She continued with her PhD on hadron production in $\pi$-deuterium interactions with a pion beam energy of 360\,GeV at Fermilab, in the 15 foot bubble chamber~\cite{Moriyasu:1978jm}. 
She was studying double scattering on deuterium, which is the scattering of the incident beam on both the neutron and proton in the deuterium target, which was established, and learned the concept of Monte Carlo simulations. Still the understanding of dynamics of the strong interaction was missing in Monte Carlo programs. Her office was in a cellar close to a recently installed terminal to an upgraded version of a CDC 6600 computer, located some 30 km away, a console reading the cards. Many people were passing by that floor and talking to Halina, so that the social atmosphere was great.  There was much more time spent on thinking and waiting for results at that time, punching cards, producing histograms and having all the time to study them, until the next step of the analysis. PAW, when everything could be done basically online, was still far away. 

At that time there was not a lot of understanding of the dynamics  of particles production in hadron interactions, apart from the fact that it was clear that they were not distributed randomly in phase space. Halina had studied Regge theory at University but she was also aware of  QCD which was starting at that time, the $J/\Psi$ had been discovered, the parton model had been developed, and people were looking for jets. For her postdoc, however, she wanted to move  from bubble chambers to more modern counting experiments. She then got a position as CERN paid associate where she joined the group of Jack Steinberger on the CDHSW (CERN-Dortmund-Heidelberg-Saclay-Warsaw) neutrino experiment.

\section{Hunting bugs at CERN}

Halina then started working on deep inelastic scattering (DIS) and this is the period when her career took off. With the construction of the SPS, higher energy neutrino beams became possible, accompanied by new more powerful detectors. The CDHSW detector was composed of target, hadron scintillator calorimeter  with iron plates immersed in a toroidal magnetic field as absorber, and muon spectrometer integrated in 19 modules. Between modules there were drift chambers for track reconstruction. The first task of Halina was the determination of the charged current cross-section with higher intensity wide-band neutrino beam, where she learned the concept of unfolding. The main measurement of the experiment was the structure function $F_2$ and, looking for an original subject, Halina tried to extract the gluon distribution from the derivative of $F_2$, assuming the DGLAP splitting functions. Too many assumptions had to be made for such an extraction, so that at the end the work was never published. However, she was by now well known as somebody thorough with the data, and joined an effort to measure the Weinberg angle $\sin^2 \theta_W$, from the ratio $R_\nu$ of neutral current (NC) and charged current (CC) cross sections, which became the subject of her habilitation~\cite{Abramowicz:1986vi} (see also later result~\cite{Blondel:1989ev}).

In isoscalar targets, assuming massless $u$ and $d$ quarks, the relation between $R_\nu$ and the Weinberg angle is

\begin{equation}
    R_\nu = \frac{\sigma_{NC}}{\sigma_{CC}} = 1/2 - \sin^2 \theta_W +5/9 \sin^4 \theta_W (1 + r^{(-1)}),
\end{equation}
where $r$ is the ratio of anti-neutrino and neutrino CC cross sections.

It is in that period that she had two major achievements. Radiative corrections were calculated and had to be taken into account for the CC and NC cross sections, however they turned out to be unreasonably different for protons and neutrons. Halina took the original paper,  coded from scratch all the formulas and got the correct result. The second achievement was in the measurement of  $\sin^2 \theta_W$ itself. From the ratio of the CC and NC cross section, CDHSW obtained a very precise result at the time, 

\begin{equation}
\sin^2 \theta_W = 0.225 \pm 0.005 \mathrm{(exp)} \pm 0.003 \mathrm{(theo)} +0.013 (m_c-1.5~\mathrm{GeV/c^2}).     
\end{equation}

The last term is due to the correction for the charm mass $m_c$, which affects the suppression of production of charm quark in CC events near threshold. Assuming an uncertainty of 0.3 GeV on $m_c$, the theoretical uncertainty increased to 0.005.   Also the CHARM experiment was measuring the Weinberg angle and in a big colloquium both experiments presented their $\sin^2 \theta_W$ measurement with around 2\% precision, the best achieved precision at the time. However, the two  reported values were 5$\sigma$ away from each other. Trying to understand the CHARM results step-by-step, Halina realized that their charm correction was applied in the wrong direction. She was told at that occasion "you are worth all the money" as the CDHSW result came out  strengthened. Her role and history at CERN in that period was that to chase and fix bugs. 

There were two women in the group at the time. CERN in the 80s was great. There were many experiments and people and you would go to the cantine for coffee and learn new things. Just before the start of LEP, and with many results from existing experiments at that time, there was overall excitement and Halina made many friends in that period.

\section{Intrigued by the proton structure at DESY}

Following the choice in picking interesting topics and pursuing her interest in the strong interactions, Halina  had still a lot of questions about proton structure functions. In 1986 she then visited DESY together with Roman Walczak, where the ep collider HERA was under construction,  and had the first contacts with G\"unter Wolf. They had the intention to contribute as Warsaw University with a component to the ZEUS detector, component which later became the Backing Calorimeter. The central part of the ZEUS detector consisted in  a vertex (later microvertex) detector, a central tracking chamber inserted in the magnetic field of a thin superconducting coil, an Uranium-scintillator electromagnetic and hadronic calorimeter surrounding the coil over the full solid angle, the backing calorimeter (BAC), and barrel as well as rear and forward muon chambers. The BAC was responsible for catching the end of the hadronic showers and detecting the muons in the bottom layer, where no muon chambers were present.

\begin{figure}
    \centering
    \vspace{-3cm}
    \includegraphics[width=0.7\textwidth]{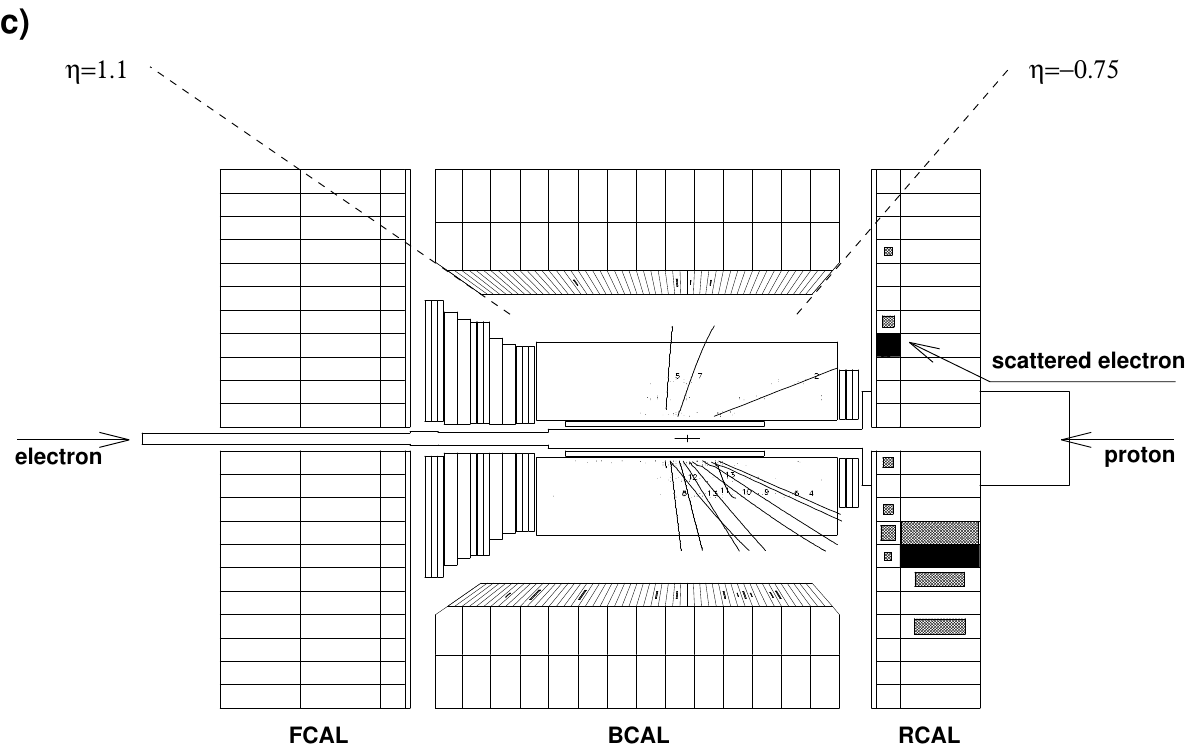}
    \includegraphics[width=0.7\textwidth]{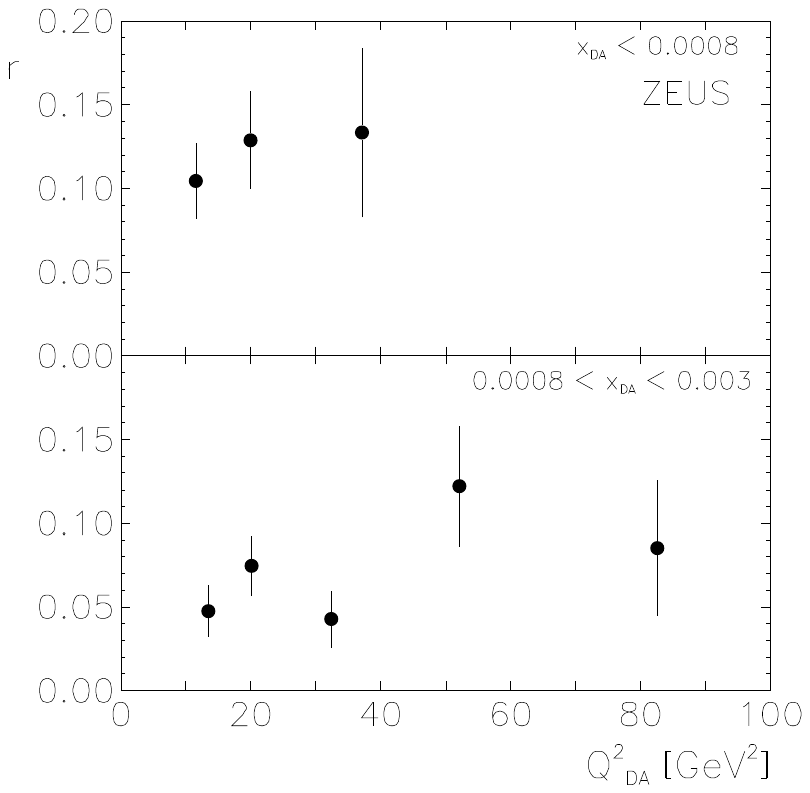}
    \vspace{-3cm}
     \caption{An event display of an LRG event at ZEUS, where the gap and absence of calorimeter clusters in the forward region is evident. The ratio of LRG events over DIS events versus the $Q^2$ variable, in two $x$ region (from~\cite{ZEUS:1993vio}). }
     \label{fig:etamax_ZEUS}
\end{figure}

With the start of data-taking she took a tenure-track position at Tel Aviv and dedicated herself to data analysis, thanks to her previous experience on structure functions (SF). She declined the position of exotics convener, while Allen Caldwell, appointed as SF convener, asked her to be his deputy. It was the beginning of a great friendship and collaboration over many years. ZEUS was second in measuring the first $F_2$, as the H1 Collaboration was the first to present at the DIS93 workshop in Durham the steep rise of $F_2$ at low-$x$~\cite{H1:1993jmo}. However, by analyzing DIS NC events, the groups were observing events without energy in the very forward region of the forward part of the calorimeter (FCAL), in the direction of the outgoing proton, and discarding these events as an anomaly (Fig.~\ref{fig:etamax_ZEUS}). The fact that this type of events did not exist in  the DIS MC simulation became a real issue.

Allen Caldwell was intrigued by these anomalous events and formed a group, including Halina, to look into them. She scanned all of the events and she did not find anything wrong, in the sense that they were well reconstructed DIS events, with a scattered electron and a perfect jet from the struck quark.

\begin{figure}
    \centering
\includegraphics[width=0.3\textwidth]{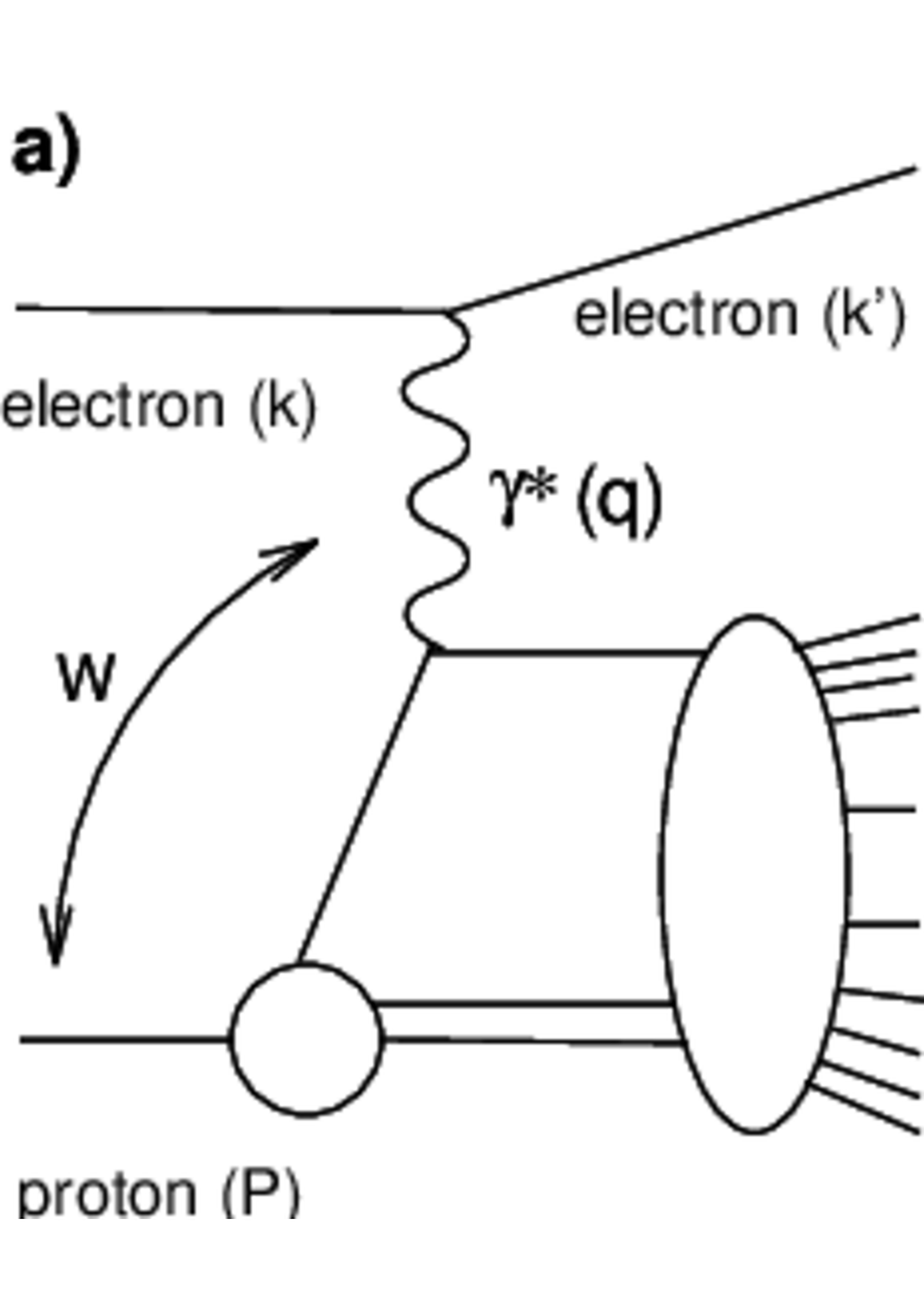}
\includegraphics[width=0.3\textwidth]{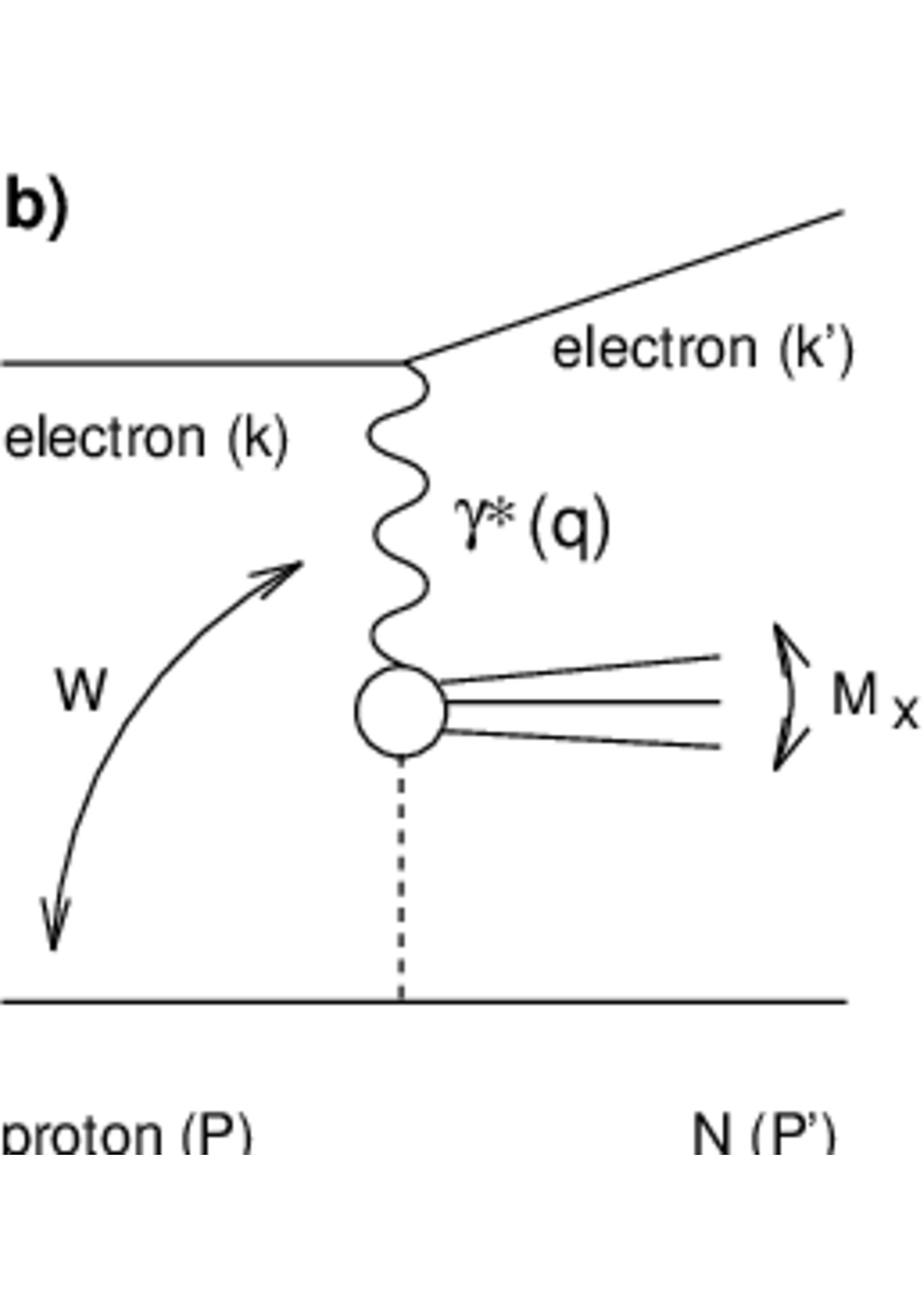}
        \caption{A DIS event at HERA at high $Q^2$ and a diffractive event (from~\cite{ZEUS:1993vio}). } 
        \label{fig:LRG_ZEUS}
    \end{figure}

With her background in Regge theory, she asked Aharon Levy if these could be diffractive events and Aharon Levy suggested to look at the hadronic mass distribution $M_X$, that should have followed a behaviour like $1/M^2_X$. By plotting the $M_X$ distribution for those events, Halina  found out that indeed it was following that behaviour. Allen Caldwell, Lothar Bauerdick, Stefan Schlenstedt and Halina invented the $\eta_{max}$ variable to select this type of events, which were later called large rapidity gap (LRG) events. The variable was  defined as the maximum pseudorapidity of all calorimeter clusters in an event (Fig.~\ref{fig:etamax_ZEUS}), and these anomalous events were concentrated at low values of $\eta_{max}$, with a large gap towards the forward (proton-remnant) region. The large excess of LRG was not explained by the DIS Monte Carlo simulation. Her first presentation at the analysis group showed that the ratio of LRG events with respect to the total number of DIS events was a constant versus $Q^2$, which was an unexpected result, as it showed that it was a leading twist behaviour. In normal DIS events, there is a colour flow transfer between the struck quark and the proton remnant. These LRG events were suggesting a colourless object, like the Regge pomeron, emitted from the proton  and interacting with the electron. The fact that these events were a constant fraction with $Q^2$  opened a new field of diffraction at high $Q^2$, with the possibility to study the pomeron structure with a virtual photon (Figs. ~\ref {fig:etamax_ZEUS}, ~\ref{fig:LRG_ZEUS}). The discovery was published in~\cite{ZEUS:1993vio}.

The atmosphere in ZEUS was great.  There were not many senior women in the Collaboration, one could count four of them, however very visible and active. The working groups were working well, meeting regularly once a week at DESY, and people were collaborating closely together. Creating instead groups working on tools, like we have now in the large LHC Collaborations, did not work out at that time,  as each group was preparing the tools that best suited their analysis. However, Halina developed with students two very innovative important tools, that were ahead of the time, and were the basis of all ZEUS analysis for the next two decades.

The reconstruction of low-$x$, low-$Q^2$ events was a challenge at ZEUS at that time, as the scattered jet was in the rear direction (RCAL) and could overlap with the scattered electron. Allen Caldwell and the Columbia group developed the first "island" calorimeter clustering algorithm. Based on that principle,  Halina and her PhD student Gennady Briskin came with the idea to match the calorimeter clusters with tracks~\cite{ZEUS:1998rvb}, improving their resolution at low transverse momenta. The reconstruction of these objects, called later "ZUFOs", is basically what today we would call Particle Flow, but this was developed already in the early 90s. The resolution in the kinematic variable $y$ improved considerably.
The second novelty tool was the development of an "electron finder", so an algorithm to find the scattered electron, using a neural network, together with the PhD student Ralph Sinkus~\cite{Abramowicz:1995zi}. They realized that they needed more variables for an electron algorithm to distinguish it from pions, and  in particular those related to the showering properties in the Uranium-scintillator calorimeter. Ralph Sinkus got interested in neural networks and proposed to use them, even if they were not very commonly used at that time in high-energy physics. The SINISTRA finder remained the best electron algorithm in ZEUS at low $Q^2$ for the next two decades~\cite{Sinkus:1996ch}.

Halina got her tenure-track position at Tel Aviv University in 1993, when the LHC endeavor was starting, and her task was to insert Tel Aviv in the LHC program. At the end, the institutes from Israel joined in a common contribution to the forward muon trigger of the ATLAS experiment. Halina, however,  was not excited at that time about the LHC program. She was instead still attracted by HERA physics, which had just started and had just pinned down the grow of the gluon parton density at low-x, so that she continued in ZEUS, where she was also physics coordinator in 2007.
She was not involved in ATLAS until she was asked to be deputy and then chair of the ATLAS publication committee (2013-2015), and started also an analysis on double parton scattering, using four-jets and later four-leptons events~\cite{ATLAS:2016rnd,ATLAS:2018zbr}. During the COVID pandemic, she initiated together with an Indian postdoc a study of double-parton scattering in two-photon final states. It looked very promising. Unfortunately, it turned out that uncertainties in the knowledge of single-parton production of two photons make the determination of the double-parton contribution unreliable and the study ended without publication after the funding run out. Till today, she is wondering whether there was a better way to attack the problem.

\section{Looking into the future}

While still involved in ZEUS and ATLAS, Halina got involved in future $e^+ e^-$ colliders, where she wanted to take the responsibility at Tel Aviv of a small but crucial component. Luminosity measurement, as vital component for all detectors and involving a precision measurement appealed to her as the next project. Achim Stahl was the initiator of the FCAL project~\cite{Stahl:2004xx,Stahl:2005xv}
while he was at DESY Zeuthen and the Tel Aviv group joined the project. There was a  world-wide effort on the linear collider at the time, and even if the ILC in Japan was not realized in the expected timescale, a strong wish for an electron-positron collider as next facility raised from those studies. Halina was always supportive of a Higgs factory, as she sees the  Higgs boson as a {\em garbage collector}, coupling to all SM particles, so that precision measurements of its couplings are a safe bet to pin down new physics. While the European community agreed on a future Higgs factory, it got split into the linear and circular collider options. Halina always supported the linear collider option as the best one for the future, for its possibility to be extended to reach higher center-of-mass energies. The circular option is attractive for the possibility to convert into a future proton-proton collider at $\sqrt{s}$ around 100 TeV. 

\begin{figure}
    \centering
\includegraphics[width=0.9\textwidth]{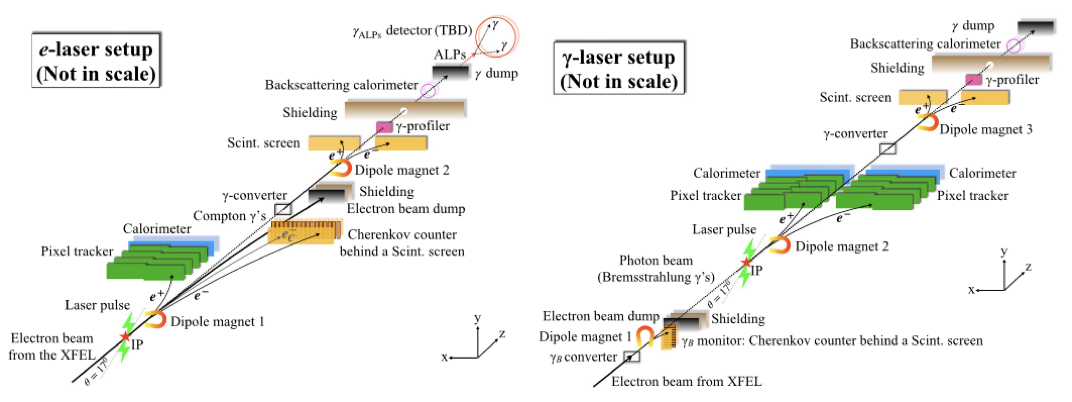}
        \caption{Schematic view of LUXE in the two different run configurations. The experiment can run colliding the electron beam directly with the laser ($e$-laser setup), or convert the electron beam in a photon beam and make this collide with the laser ($\gamma-$laser setup. (from~\cite{Abramowicz:2021zja}). } 
        \label{fig:LUXE}
    \end{figure}

It is in this period that Halina was elected ECFA chair for three years and then appointed as secretary of the European Strategy update for Particle Physics, covering the period 2014-2020~\cite{EPPSU2020}. These were six years when she  dedicated herself to scientific management, however still learning a lot.  She was the first woman as ECFA chair and at the question if she thinks she brought anything different to the committee as a woman, Halina humbly answered that one should ask the committee members. She has been always very determined in her scientific management roles, but also always payed attention that each member contributed and that the final report was delivered. As a daughter of a diplomat and growing up in an international environment, she had learned since her youth to interact with people. Around 300 papers were submitted as input to the European Strategy update of 2020, covering a wide spectrum of topics.  The community was divided at the time in a linear (ILC, CLIC,..) collider option and in the circular FCC option, and strong opinions were expressed on both sides during the process, which Halina had to mediate for the final document. Halina recently expressed her opinion that a project like FCC is too expensive and does not leave space for other initiatives at CERN for many decades and that the HEP community should better invest in driving novel accelerator technologies. Higher energies could then be reached with much easier infrastructure, also cheaper to maintain. She supports the P5 conclusion on recommending~\cite{P52023} a 10 TeV parton-parton machine and their encouragement of novel technology. The European strategy update is a hot topics these days as the next update process is going on these months.

Being close to retirement, Halina looked for another interesting project, smaller compared to the LHC experiment scale,  and she was attracted by the LUXE experiment at DESY, started by Beate Heinemann, Andreas Ringwald and Reinhard Brinkmann. The LUXE (Laser Und XFEL Experiment) experiment at DESY/EuXFEL, Hamburg, combines the high-quality, high-energy electron beam of the European XFEL with a powerful laser (Fig.~\ref{fig:LUXE}),  to measure QED in a strong non-perturbative regime~\cite{Abramowicz:2021zja}. In the Schwinger limit $\epsilon_{cr}=m_e^2 c^3 / (e \hbar)=1.32 \times 10^{18} V/m $, strong-field vacuum polarization effects give rise to non-linearities on the properties of the vacuum, like the creation of real electron-positron pairs from the vacuum. While QED is more part of the SM theory, strong-field QED is more in the field of plasma and laser physics, so being very interdisciplinary. Halina thought that as high-energy  physicist, she could add a lot to the experiment, like building redundancy in the detectors to be able to control systematics and understand correlations in the measurements, and joined the calorimetry with the Tel Aviv group. Due to the difficulties in funding, the experiment got delayed compared to its original schedule, but recently substantial progress has been made in the funding of the beamline (ELBEX EU infrastructure grant awarded in 2024) and borrowing a laser (JETI40 loaned by Jena University), so that the experiment could start in 2030.

As many of us in the same generation, Halina is not able anymore to do data analysis herself, but she has still a lot of ideas.  Her future now is LUXE, a project that she really would like to succeed. An advice for the younger generations? Always choose projects that are exciting and interesting for you, that motivate you in writing grants and getting to work in the morning.

\section{Acknowledgements}

Prof. Halina Abramowicz is full professor at the University of Tel Aviv. She is presently member of HEPAP, chair of the Physics Advisory Committee of Fermilab (USA), member of the Scientific Council of the DESY laboratory (Germany), and chair of the Advisory Board of the ORIGINS Cluster of Excellence, Munich (Germany). Recently, she was also invited to be a member of the Scientific Board of the Enrico Fermi Research Centre. She has been awarded the 2021 Beate Naroska
award from the Cluster of Excellence Quantum Universe of the University of Hamburg and DESY, Germany. This award has inspired this short paper, which is supported by the Deutsche Forschungsgemeinschaft (DFG, German Research Foundation) under Germany’s Excellence Strategy – EXC 2121 „Quantum Universe“ – 390833306.

\printbibliography

\end{document}